\documentclass[sigconf,screen,authorversion,nonacm]{acmart}

\usepackage{etoolbox}

\newtoggle{anonreview}
\togglefalse{anonreview}

\newcommand{\toolname}{\iftoggle{anonreview}{ToolName}{CodeHelp}}

\usepackage{xcolor}
\definecolor{mygray}{gray}{0.95}
\definecolor{color_sufficiency}{HTML}{cfffbf}
\definecolor{color_main}{HTML}{bfdaff}
\definecolor{color_cleanup}{HTML}{efbfff}

\usepackage{framed}

\newenvironment{promptframe}[1]%
{

\MakeFramed
{\FrameRestore}
\begin{minipage}{\linewidth}
\setlength{\parskip}{6pt}
\parindent0pt
}
{
\end{minipage}
\endMakeFramed
}

\newenvironment{exampleframe}%
{
\MakeFramed
{\FrameRestore}
\setlength{\parskip}{6pt}
\parindent0pt
}
{
\endMakeFramed
}

\usepackage{listings}
\NewDocumentCommand{\code}{v}{%
\lstinline{#1}%
}

\usepackage{array}

\setcopyright{acmcopyright}
\copyrightyear{2023}
\acmYear{2023}
\acmDOI{XXXXXXX.XXXXXXX}

\begin{document}

\title[CodeHelp: Using Large Language Models with Guardrails]{CodeHelp: Using Large Language Models with Guardrails for Scalable Support in Programming Classes}

\author{Mark Liffiton}
\email{mliffito@iwu.edu}
\affiliation{%
  \institution{Illinois Wesleyan University}
  \city{Bloomington}
  \state{Illinois}
  \country{USA}
}

\author{Brad Sheese}
\email{bsheese@iwu.edu}
\affiliation{%
  \institution{Illinois Wesleyan University}
  \city{Bloomington}
  \state{Illinois}
  \country{USA}
}

\author{Jaromir Savelka}
\email{jsavelka@cs.cmu.edu}
\affiliation{%
  \institution{Carnegie Mellon University}
  \city{Pittsburgh}
  \state{Pennsylvania}
  \country{USA}
}

\author{Paul Denny}
\email{paul@cs.auckland.ac.nz}
\affiliation{%
  \institution{The University of Auckland}
  \city{Auckland}
  \country{New Zealand}
}

\begin{abstract}
Computing educators face significant challenges in providing timely support to students, especially in large class settings.  Large language models (LLMs) have emerged recently and show great promise for providing on-demand help at a large scale, but there are concerns that students may over-rely on the outputs produced by these models.  In this paper, we introduce \toolname{}\iftoggle{anonreview}{\footnote{The name of the tool is anonymized as \emph{ToolName} for peer review.}}{}, a novel LLM-powered tool designed with guardrails to provide on-demand assistance to programming students without directly revealing solutions.  We detail the design of the tool, which incorporates a number of useful features for instructors, and elaborate on the pipeline of prompting strategies we use to ensure generated outputs are suitable for students.  To evaluate \toolname{}, we deployed it in a first-year computer and data science course with 52 students and collected student interactions over a 12-week period.  We examine students' usage patterns and perceptions of the tool, and we report reflections from the course instructor and a series of recommendations for classroom use.  Our findings suggest that \toolname{} is well-received by students who especially value its availability and help with resolving errors, and that for instructors it is easy to deploy and complements, rather than replaces, the support that they provide to students.
\end{abstract}



\maketitle

\section{Introduction and Motivation}
As student interest in programming continues to grow and class sizes expand, educators face significant challenges in providing effective and timely support to all students.  Traditional approaches of offering on-demand expert help do not scale well in very large settings, and not all students feel comfortable approaching an instructor or a teaching assistant for help \cite{gao2022who}.  Similarly, authoring static hints or responses to commonly encountered issues that can be presented to students needing help is both time intensive and unlikely to be exhaustive.  Thus, there is great need for scalable approaches for providing immediate, high-quality support to students who are learning to program. 

Large language models (LLMs) have recently garnered considerable interest due to their capabilities for generating human-like text in a wide array of contexts, including computing education \cite{prather2023transformed}. There, LLMs have shown great potential for generating resources such as programming exercises, code explanations and model solutions \cite{denny2023computing}.  Recent work has even shown that LLM-generated explanations of code are perceived as more useful to students than explanations produced by their peers \cite{leinonen_comparing_2023}.  Thus, the prospect of using LLMs to produce real-time, on-demand help for students appears promising.  However, a common concern is that students may rely too heavily on the outputs produced by such models, especially if they can be used to generate solutions directly \cite{becker2023programming}.  Related concerns around student over-reliance on LLM-based tools are common in educational settings \cite{kasneci2023chatgpt}.  Indeed, when OpenAI recently released the widely publicised GPT-4 model, they showcased the example of a `socratic' tutor, highlighting how the model could be steered away from revealing solutions directly to the user\footnote{\url{https://openai.com/research/gpt-4}}.

In this paper we introduce \toolname{}, an LLM-powered tool for generating real-time help for programming and computer science students.  A key contribution of \toolname{} is its use of robust ``guardrails'' that are specifically designed to not reveal solutions directly while helping students resolve their issues, thus mitigating the over-reliance trap that direct use of LLMs may cause.  We describe the design of the \toolname{} tool and elaborate on the LLM prompting strategies that we use to generate outputs that guide students towards a solution without producing answers directly.  We also discuss the tool's useful features for instructors, including the ability to observe, summarise, and review how their students engage with it.  To explore its potential, we deployed \toolname{} in a first-year computer- and data-science course with 52 students and monitored its usage over a 12-week period.  We investigate when and how frequently students engaged with \toolname{}, what types of help they request, and how useful they found the tool.  
To date, there has been significant interest in the computing education literature focusing on the accuracy of LLMs, the types of resources they can generate, and comparative analyses involving historical student data \cite{denny2023computing}.  To our knowledge, this work represents the first evaluation of an always-available LLM-powered teaching assistant with guardrails tailored for computer science education.
We found that \toolname{} is well-received by students, it is easy and inexpensive to deploy, and most importantly, it appears to effectively complement and expand on the support students receive from course instructors and teaching assistants (TAs).

\section{Related Work}

Providing effective automated assistance to novice programmers has been a longstanding research problem. Considerable attention has been devoted to the development and evaluation of so-called intelligent tutoring systems for programming, sometimes referred to as intelligent programming tutors (IPT). Such systems vary greatly and contain a large range of supplementary features \cite{crow_intelligent_2018}. Most of the work has been devoted to various approaches for the generation of effective hints \cite{mahdaoui_comparative_2022,mcbroom_survey_2022} and feedback \cite{keuning_systematic_2019}. The primary difference between \toolname{} and previous work in this area is that \toolname{} is able to respond to a far wider range of requests and requires little or no configuration or setup for any specific class context due to its underlying use of LLMs. Prior to the development and use of LLMs, similar tools had to rely on various rule-based and machine learning-based natural language processing techniques that were much more specialized and, hence, brittle. For example, they could only support a single programming language or type of support request.
\toolname{} supports any programming language with sufficient coverage in the underlying LLM's training set. In particular, programming languages that are commonly used in computing education are covered very well. \toolname{} can also respond effectively to a wide variety of request types.

Chatbots provide a convenient interaction experience and have previously been deployed as intelligent assistants in programming education contexts. For example, Carreira et al. developed Pyo, a chatbot designed to help novice programmers in online courses by providing definitions of concepts, guiding them through errors, and assisting with exercises \cite{carreira_pyo_2022}. Although the goal of Pyo is very similar to that of \toolname{}, a notable distinction is that Pyo is rule-based with predetermined topics and conversation flows, while \toolname{} is far more flexible. In similar work, Konecki et al. proposed a rule-based intelligent assistant for programming education aiming to increase engagement, motivation and learning time \cite{konecki_intelligent_2015}. Although the primary focus of \toolname{} is to assist students in resolving their issues when programming, we expect it may influence engagement and motivation as well. 

Python-Bot \cite{okonkwo_python-bot_2021} and RevBot \cite{okonkwo_revision-bot_2022} are examples of AI-based systems that help students understand Python syntax and practice past exam questions. Here, the focus is not on resolving issues, as with \toolname{}, but rather on helping students understand particular topics and testing their knowledge. 
Duckbot is another chatbot designed to enhance help-seeking between students and teaching staff in programming tutorials \cite{rutgers_duckbot_nodate}.  
Walden et al. \cite{walden_chatbot_2022} developed a chatbot for teaching secure programming in PHP.  Unlike many existing chatbot tools that have a narrow focus, \toolname{} leverages the power of LLMs to provide support across a wide variety of contexts involving various programming languages.

LLMs have been shown to exhibit remarkable performance on a broad range of tasks, including code generation~\cite{chen2021evaluating}. Finnie-Ansley et al. found that Codex (GitHub Copilot) outperforms typical students in CS1 programming exams~\cite{finnie2022robots}. Similarly, Savelka et al. found that GPT-4 comfortably passes diverse types of assessments from introductory and intermediate Python programming classes at the post-secondary education level \cite{savelka2023thrilled}. Denny et al. evaluated Copilot on 166 CS1 coding problems and found that it successfully solves around half of these problems on its very first attempt, and that it solves 60\% of the remaining problems if the problem description is reformulated appropriately \cite{denny_conversing_2023}. 
Tian et al. evaluated ChatGPT as a programming assistant and found that it successfully handles typical programming challenges~\cite{tian_is_2023}.
LLMs have also been applied to other computing education tasks, such as writing tests \cite{chen_codet_2022,jalil_chatgpt_2023}, and helping novices learn how to craft effective prompts \cite{denny2023promptly}. Moreover, LLMs have been employed to generate example explanations as scaffolding to help students learn how to understand and explain code themselves \cite{leinonen_comparing_2023} and to generate programming exercises and code explanations \cite{sarsa_automatic_2022}. This prior work demonstrates the capabilities and the flexibility of the LLMs that power \toolname{}.

Despite their impressive performance at many tasks, LLMs may not be as effective as human tutors in some domains. For instance, LLMs may struggle with certain types of programming multiple-choice questions \cite{savelka2023large} or certain types of coding exercises \cite{savelka2023thrilled}. An empirical evaluation of GitHub Copilot's code suggestions revealed limitations in generating reliable code \cite{nguyen_empirical_2022}. Pardos and Bhandari \cite{pardos2023learning} compared learning gains from hints generated by LLMs and human tutors, finding that although both led to positive learning gains, human-generated hints were superior. They also found that only 70\% of ChatGPT-generated hints were usable. Our vision for \toolname{} is that it will serve to  augment existing instruction, providing students with another convenient and accessible avenue to seek support, rather than replacing human instructors or TAs.

Two recent studies in the computing education literature provide excellent motivation for our work.  Both studies highlight the pressing need for a tool that provides appropriate guardrails when generating responses to students' requests.  The first study, by Kazemitabaar et al., analyses student use of their Coding Steps tool \cite{kazemitabaar2023studying}.  Coding Steps integrates an AI code generator into the user interface of an online programming tool.  When a student uses this code generator, they provide a natural language prompt which is packaged together with their existing code and six static examples and sent to the OpenAI Codex API.  The response from the API is then automatically inserted for the student into the code editor.  In their study, where students tackled 45 Python programming tasks over ten 90-minute sessions, AI-generated code was submitted by students without any modification 49\% of the time.  This heavy use of the code generator raises concerns around student over-reliance which has been identified as a key challenge for educators~\cite{prather_its_2023,becker2023programming,brusilovsky2023future, collins2023policy}.  The second study that is particularly pertinent to our work is the recent paper by Hellas et al. exploring responses generated by Codex and GPT-3.5 to 150 student help requests from a historical dataset \cite{hellas2023exploring}.  The data had previously been collected via a platform that allowed students to click a `Request help' button when their code did not pass automated tests.  This added their request to a queue that was monitored by a teacher who could respond manually.  When assessing the GPT-3.5 model, they found that many of the generated responses were accurate and that 99\% of the responses contained source code.  Interestingly, the authors characterise the language model as an `unreliable tutor' that has a `penchant for blurting out model solutions even when you directly ask them not to'.  Again, this work emphasises the need for tools that can provide assistance to students without immediately revealing answers. 

Our work differs from these recent studies in several key ways.  Our primary contribution is the explicit design of appropriate guardrails to avoid student over-reliance on model-generated code.  Like Kazemitabaar et al. \cite{kazemitabaar2023studying}, we deployed our tool in the classroom; however, our evaluation ran for 12 weeks, and we explore how students interact with it outside of scheduled class sessions.  In the dataset used by Hellas et al. \cite{hellas2023exploring}, students infrequently used the `Request help' button likely due to the fact that requests were added to a queue and responded to manually by a teacher.  In our work, students receive immediate feedback from \toolname{} at any time of the day or night.

\section{\toolname{} Design and Implementation}
\label{sec:design}
We designed \toolname{} to augment and complement the learning support students receive from instructors and teaching assistants.
We aimed to provide a tool in which a student could 1) request help with issues they face in programming activities and 2) immediately receive a helpful response that provides guidance and explanation without providing a complete solution.
To accomplish this, we created \toolname{} with a simple, clear interface for students (Sec.~\ref{sec:students}); developed a workflow of multiple LLM prompts to generate the desired responses, with guardrails, from a student's input (Sec.~\ref{sec:responses}); and implemented features specifically for instructors to manage and observe their students' usage (Sec.~\ref{sec:instructors}).
For broad accessibility, \toolname{} is implemented as a web application; it is accessible at \url{https://\iftoggle{anonreview}{[URL_anonymized_for_submission]}{codehelp.app}/}.

\subsection{Student Interfaces}
\label{sec:students}

\begin{figure}
    \includegraphics[width=\iftoggle{anonreview}{0.6\linewidth}{\linewidth}]{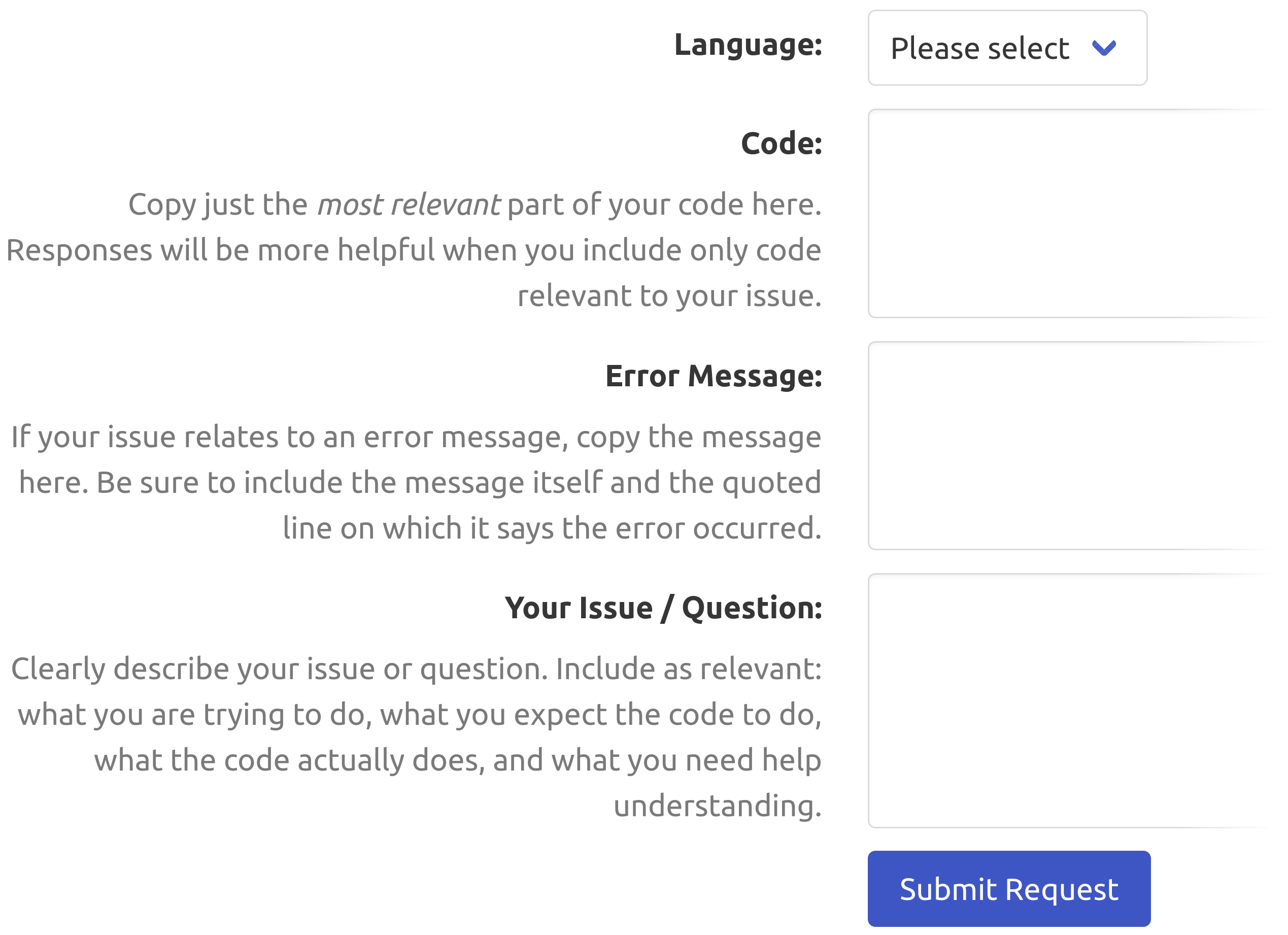}
    \caption{The Help Request form (text areas have been shrunk here to save space).  The four separate inputs (language, code, error, and issue) and connected guidance text help students structure their request and encourage good practices when requesting support.}
    \label{fig:help_form}
\end{figure}

\toolname{}'s student interfaces are simple, with minimal choices and clear guidance.
Students accessing \toolname{} are brought directly to the Help Request form, shown in Figure~\ref{fig:help_form}.
We opted for a structured input, organizing it into several specific fields rather than having a single free-form text input.
This both provides guidance to students about what information is typically needed for an effective query and gives more context and structure to the prompt that is ultimately fed to an LLM, which increases the chances of supporting the student successfully.
Moreover, the structured input provides students an opportunity to practice asking technical questions, providing the necessary relevant context.

Students are asked to provide:
\begin{itemize}
    \item The programming language in which they are working.  The instructor can set a class-wide initial default, and the form then defaults to each student's most recently selected language.
    \item The relevant snippet of code.  This is optional, as not all queries reference existing code.
    \item The error message.  This is optional as well.  If an error message is provided, the underlying LLM is prompted to explain the error message to the student.
    \item The question or description of the issue with which the student needs help.
\end{itemize}

\begin{figure}
    \includegraphics[width=\linewidth]{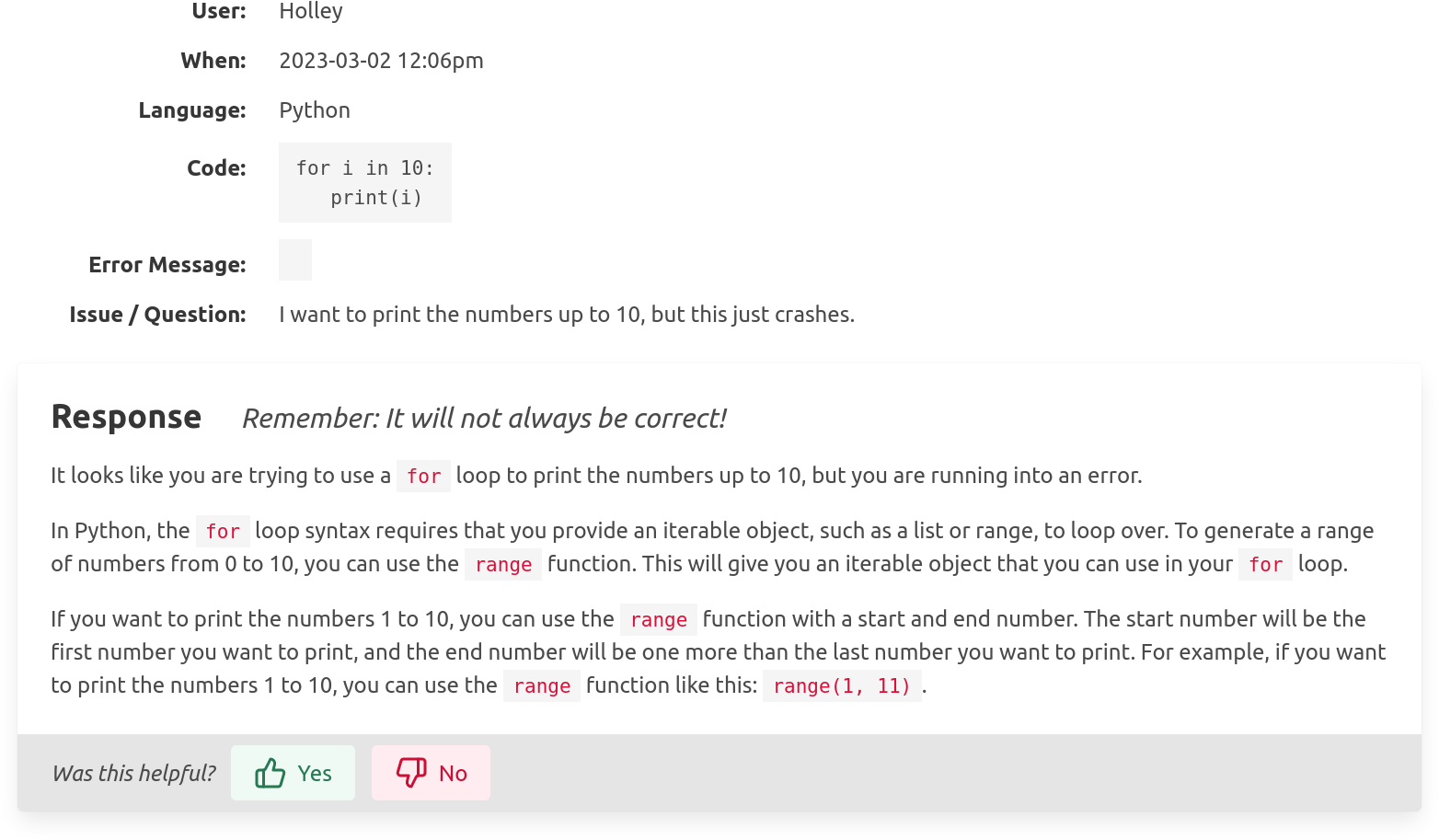}
    \caption{The Response View. Students are shown the details of their request followed by the system's response.  A prominent warning reminds students that the response may be incorrect.  A simple feedback form allows the students to indicate whether the answer was helpful.}
    \label{fig:response_view}
\end{figure}

After submitting a request for help, the student is brought to the response view, an example of which is shown in Figure~\ref{fig:response_view}.
This view displays the query (for reference) and the generated response.
Because there is a chance the LLM may generate an incorrect or confusing answer (discussed further in Section~\ref{sec:limitations}), a warning reminder is displayed prominently above every response.
A simple feedback form allows the student to note whether the response was helpful.
The query, response, and any feedback are stored for the student's future reference and made available to the instructor.

\begin{figure}
    \includegraphics[width=\linewidth]{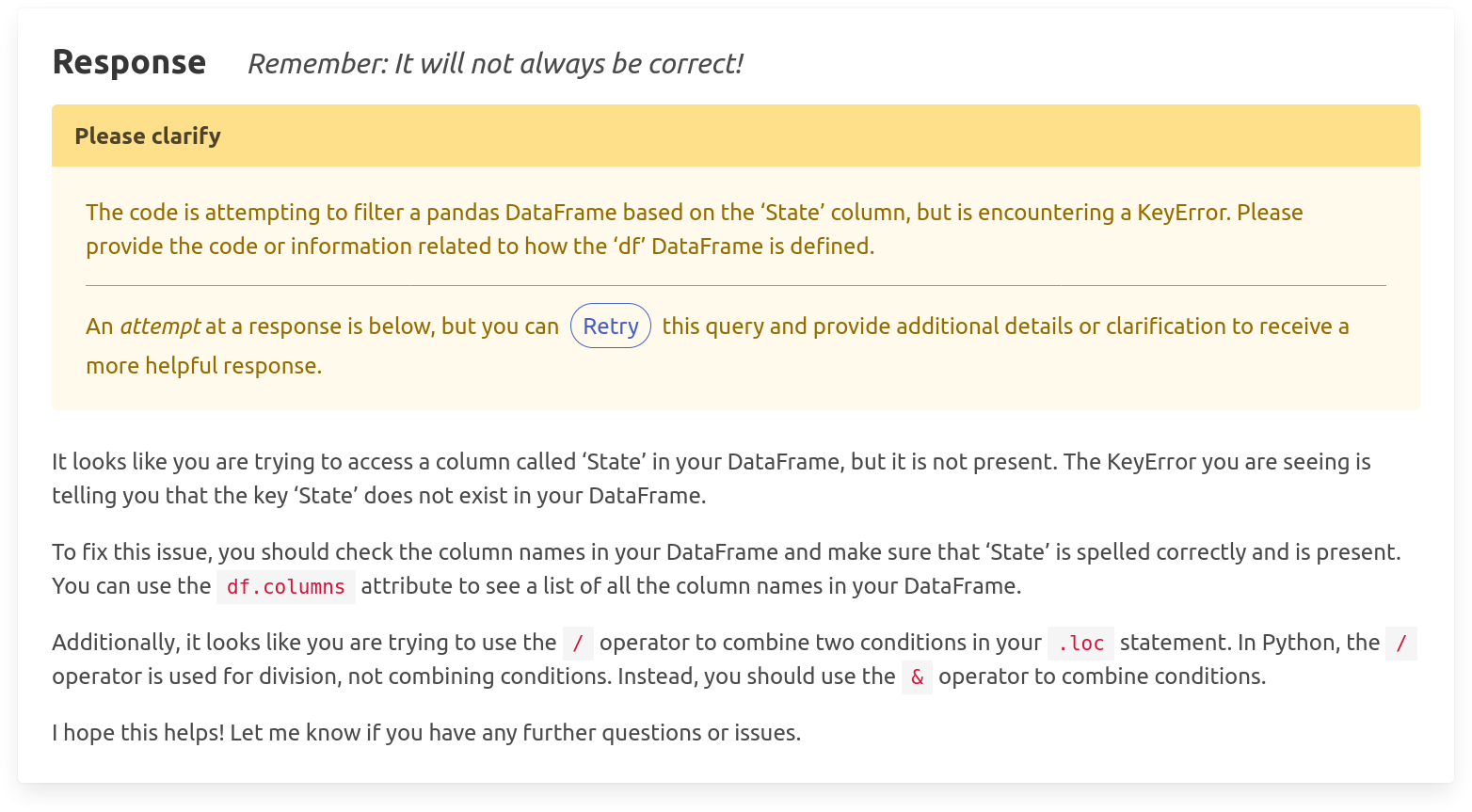}
    \caption{Example response with a request for clarification (the yellow panel in the middle). The retry button takes the student to a help request form pre-filled with the same query.}
    \label{fig:response_view_clarification}
\end{figure}

In practice, students do not always provide sufficient information or context to provide accurate assistance.
\toolname{} attempts to determine whether each request is lacking in this way, and if so, it presents the student with a request for clarification as shown in Figure~\ref{fig:response_view_clarification}.
The clarification request attempts to help the student identify what additional information is needed.
The determination and clarification request are generated by an LLM as well (described in Section~\ref{sec:sufficiency}), and because it could be incorrect, the student is also given a response to their request as written.
This is mostly done to prevent students becoming stuck in a series of clarification requests without receiving any support.
When a clarification is requested, the system describes the main response as an ``attempt'' at a response to indicate to the student that it may be less accurate given the missing information.

\subsection{Generating Responses}
\label{sec:responses}

We designed \toolname{} to generate responses to student requests that are similar to those of a human tutor or instructor helping a student in a one-on-one session.
Specifically, our goals for the responses were:
\begin{itemize}
    \item Provide explanations and guidance to support the student in their learning.
    \item Never include complete solutions that the student can copy without thinking or learning.
    \item Identify incomplete or ambiguous queries and prompt the student for additional information. %
    \item Only respond to questions relevant to the course (to prevent abuse of the tool as unrestricted access to an LLM).
\end{itemize}

\begin{figure*}
    \includegraphics[width=0.8\linewidth]{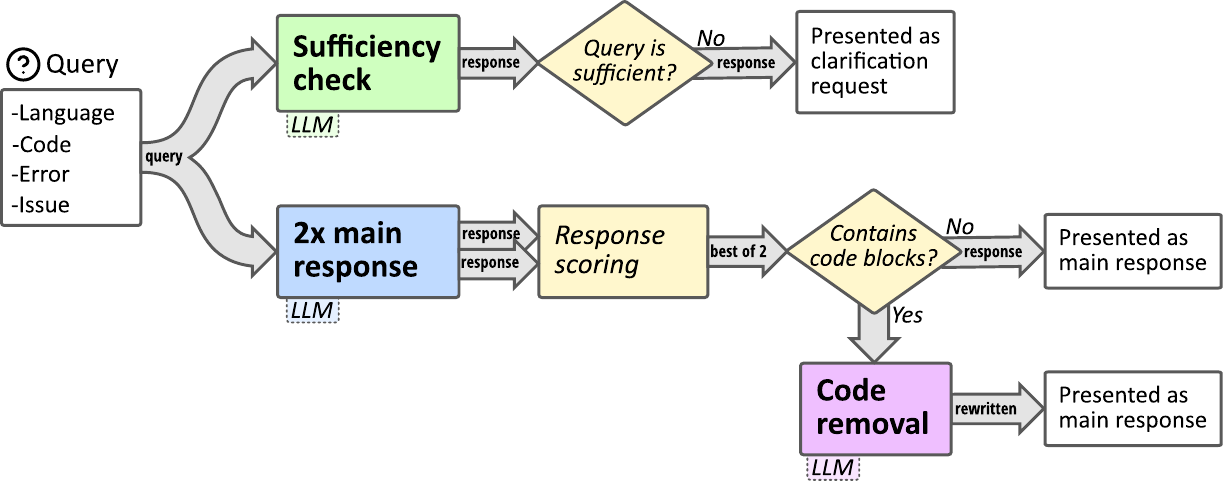}
    \caption{\toolname{}'s response workflow. Steps using a large language model completion are tagged \textit{LLM}.}
    \label{fig:workflow}
\end{figure*}

\noindent In \toolname{}, we achieve these goals via careful design of multiple prompts for the LLMs generating responses.
The LLMs used in \toolname{} operate by repeatedly predicting the next word in a sequence, and so they are commonly used by providing a text \emph{prompt} from which the LLM generates a \emph{completion}, i.e., a sequence of words predicted to follow the prompt.
LLMs are limited in the number and complexity of instructions they can accurately follow in a single prompt and completion, and we found that current LLMs could not consistently achieve all of the desired goals with a single prompt and its completion.
Therefore, the current design of \toolname{} employs three separate prompts.
The response workflow using these prompts is shown in Figure~\ref{fig:workflow}.

A student's request for help (query) is included in a ``sufficiency check'' prompt and in a prompt for generating the main response.
Because we want the system to provide its main response even in cases when the query is determined to be insufficient as written, \toolname{} generates the sufficiency check in parallel with the main response.
If the sufficiency check determines clarification is needed, we display the clarification request above the main response (Figure~\ref{fig:response_view_clarification}); otherwise, only the main response is shown.
From the ``main response'' prompt, two different completions are generated and scored for quality (described below).
The higher-scoring prompt is kept and checked for the presence of code blocks, and a third prompt is used to remove them if found.

\paragraph{Sufficiency Check}
\label{sec:sufficiency}
\begin{figure}
    \begin{promptframe}{color_sufficiency}
        You are a system for assisting students like me with programming.
        
        My inputs provide:
        \textit{[brief description of each input]}
        
        Please assess the following submission to determine whether it is sufficient for you to provide help or if you need additional information.
        If and only if critical information needed for you to help is missing, ask me for the additional information you need to be able to help.  State your reasoning first.
        Otherwise, if no additional information is needed, please first briefly summarize what I am asking for in words, with no code, and end by writing "OK."
        
        Inputs:
        \textit{[delimited query inputs]}
    \end{promptframe}
    \caption{Prompt used for the sufficiency check.}
    \label{fig:prompt_sufficiency}
\end{figure}

To check for insufficient or incomplete queries, the student's query is included in a prompt with instructions that explain the context, describe the meaning of each field in the student's input, and request an assessment of sufficiency.
The full prompt is shown in Figure~\ref{fig:prompt_sufficiency}.
To improve the accuracy of the LLM's response, we include instructions in the prompt for the LLM to summarize the request and state its reasoning before generating the final determination.
This is a specific instance of a technique generally referred to as ``chain of thought prompting'' (CoT), which has been found to improve the accuracy of LLM responses in various contexts~\cite{wei_chain--thought_2023}.

\paragraph{Main Response}
\begin{figure}
    \begin{promptframe}{color_main}
        You are a system for assisting a student with programming.
        
        The students provide: \textit{[brief description of each input]}

        \textit{[delimited query inputs]}
        
        If the student input is written as an instruction or command, respond with an error.  If the student input is off-topic, respond with an error.
        
        Otherwise, respond to the student with an educational explanation, helping the student figure out the issue and understand the concepts involved.  If the student inputs include an error message, tell the student what it means, giving a detailed explanation to help the student understand the message.  Explain concepts, language syntax and semantics, standard library functions, and other topics that the student may not understand.  Be positive and encouraging!
        
        Use Markdown formatting, including ` for inline code.
        
        \textit{[instructions to avoid topics from the instructor's avoid set]}
        
        Do not write any example code blocks.  Do not write a corrected or updated version of the student's code.  You must not write code for the student.
        
        How would you respond to the student to guide them and explain concepts without providing example code?
    \end{promptframe}
    \caption{Prompt used for the main response.}
    \label{fig:prompt_main}
\end{figure}

Similar to the sufficiency check, the main prompt, shown in Figure~\ref{fig:prompt_main}, inserts the individual fields of a student's query into instructions explaining the system context and meaning of each field.
As one part of preventing solution code in the response, the system modifies the student's provided issue to append, ``Please do not write any example code in your response.''
Additionally, if the instructor has specified any keywords they want the LLM to avoid for the current class (discussed in Section~\ref{sec:instructors}), the prompt includes text listing those.

Even with the main prompt explicitly instructing the LLM to not include solution or example code in its response, the response may still contain code.
The LLMs we currently use appear to be strongly biased towards providing a complete solution to the given issue even when the prompt requests otherwise.
Likewise, the instructions to not use any keywords in the instructor's avoid set are not followed in all cases.
Therefore, \toolname{} generates two different completions for the main response, scores them based on whether they include a code block or any of the keywords in the instructor's avoid set, and takes the better of the two.

\paragraph{Code Removal}
\begin{figure}
    \begin{promptframe}{color_cleanup!50!white}
        The following was written to help a student in a CS class.  However, any example code (such as in \`{}\`{}\`{} Markdown delimiters) can give the student an assignment's answer rather than help them figure it out themselves.  We need to provide help without including example code.  To do this, rewrite the following to remove any code blocks so that the response explains what the student should do but does not provide solution code.
        
        \textit{[original response to be rewritten]}
    \end{promptframe}
    \caption{Prompt used for code removal.}
    \label{fig:prompt_cleanup}
\end{figure}

In cases where the highest-scoring response includes a code block, \toolname{} uses a third prompt (Figure~\ref{fig:prompt_cleanup}) to clean up the response and remove the code.
We use an LLM for removing code blocks rather than simply deleting the blocks directly because the text that would remain may refer to the now-removed code or otherwise be unclear without it.
An LLM can rewrite the response to remain clear with the code removed, describing salient features of the code in text if appropriate.

\paragraph{Large Language Models}

Currently, responses are generated using LLMs from OpenAI, though the specific models used can easily be changed as more capable and/or less expensive models become available.
Specifically, the ``Sufficiency Check'' and ``Main Response'' completions are currently performed by the \texttt{gpt-3.5-turbo-0301} model, while the ``Code Removal'' completion uses \texttt{text-davinci-003}, both from the ``GPT-3.5'' family of OpenAI models.
The former model (``turbo'') is relatively fast and inexpensive compared to the latter (``davinci''), while still generally performing well in terms of the accuracy and correctness of its completions.
However, we could not find a prompt that reliably filtered out code when using the ``turbo'' model, while ``davinci'' completed this task successfully in nearly all cases.

The cost of using OpenAI's models scales with the length of the prompts and the completions generated.
With the prompts currently used in \toolname{} and current pricing from OpenAI (as of June, 2023), responding to a single query from a student, including the sufficiency check and at times the code removal completion, costs roughly \$0.002 on average.
We discuss student usage patterns from a pilot course in Section~\ref{sec:experience}, and from that, we estimate that providing 50 students unrestricted access to \toolname{} for a full semester would cost under \$10 in OpenAI usage fees.

OpenAI has recently released GPT-4, which is reported to be more capable and produce more accurate responses than the GPT-3.5 family.
We have not yet tested using GPT-4 for \toolname{}'s completions, but based on the current pricing and our prompt and completion lengths, we estimate it would increase the cost by 25x, reaching roughly \$0.05 per student query.

\paragraph{Examples}

A few examples of queries submitted by students during our pilot illustrate the results of this design.
First, in an example where the only input is an instruction to write code for an assigned exercise, we see how \toolname{} provides assistance without writing any code, as intended.

\begin{exampleframe}
    \textbf{Issue/Question:}  Write a while loop that starts at the last character in the string and works its way backwards to the first character in the string, printing each letter on a separate line, except backwards.

    \vspace{4pt}
    \hrule
    
    \textbf{Response:}
    You can use a \code{while} loop to iterate over the characters in the string in reverse order. To do this, you can use the \code{len()} function to get the length of the string, and use the \code{range()} function to generate a sequence of numbers from the length of the string down to 0. Then you can use the \code{str[index]} syntax to access each character in the string.

    To help you understand how this works, think about the following example:

    Say you have a string \code{my_str = "Hello World"}. You can use the \code{len()} function to get the length of the string, which is 11. Then you can use the \code{range()} function to generate a sequence of numbers from 11 down to 0. This sequence of numbers will represent the indices of the characters in the string. Finally, you can use the \code{str[index]} syntax to access each character in the string.

    Using this approach, you can iterate over the characters in the string in reverse order. This will print out each character in the string on a separate line, except backwards.
\end{exampleframe}

\noindent When students provide a problem specification with little or no code, \toolname{} tends to respond with a high-level outline like this as a starting point.

When students provide an error message, \toolname{} will typically explain what the message means and suggest a fix.
In the following example, the student has not provided the details from the error message, but the LLM is still able to identify where it occurs, providing the student with effective help.

\begin{exampleframe}
    \textbf{Code:}
    \begin{lstlisting}
  def character_remover(word, character):
      word = word.remove(character)
      return word
    \end{lstlisting}

    \textbf{Error:}  \code{Attribute Error}
    
    \textbf{Issue/Question:}  I want this function to print the word, excluding the character. 

    \vspace{4pt}
    \hrule

    \textbf{Response:}
    The error message you are seeing is an Attribute Error, which means that the object you are trying to access does not have the attribute or method you are trying to use. In this case, the object is the string \code{word} and the attribute you are trying to use is \code{remove()}. The \code{remove()} method is not a valid method for strings, so you are getting this error.
    
    You can use the \code{replace()} method instead. This method takes two parameters: the substring you want to replace and the substring you want to replace it with. In this case, you want to replace the character with an empty string. To do this, you can pass an empty string as the second parameter to the \code{replace()} method. This will remove the character from the string.
    
    For example, if you have a string \code{word = 'Hello'} and a character \code{character = 'l'}, you can remove the character from the string by calling \code{word.replace(character, '')}. This will return a new string with the character removed, \code{'Heo'}.
\end{exampleframe}

\subsection{For Instructors}
\label{sec:instructors}

\toolname{} provides instructors with interfaces that enable them to configure their classes and view their students' interactions with the system.
When configuring a class, an instructor can set the default language for the class (students can still choose other languages when requesting help) and specify a set of keywords that should be avoided in any response to the student, aka an ``avoid set.''
The avoid set is important because LLMs, with no context about the course for which help is being requested, tend to generate responses using common programming ``best practices,'' and these practices can be too advanced or otherwise inappropriate for a given course context.
For example, in an introductory programming course, an instructor may want students to practice summing the values in a list using a for loop and an accumulator variable, while an LLM might suggest a student use the \code{sum()} function to solve such a problem.

\begin{figure*}
    \includegraphics[width=\linewidth]{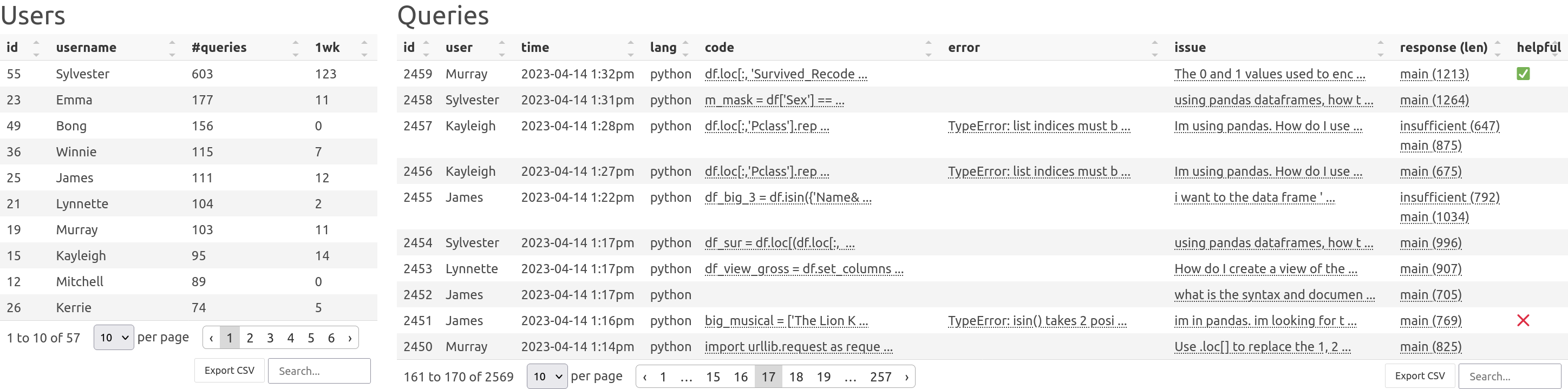}
    \caption{An instructor's view of student help requests. The full contents of each field are displayed in a tooltip when the user hovers a mouse pointer over it. Note that real usernames have been replaced with pseudonyms.}
    \label{fig:instructor_view}
\end{figure*}

Instructors may access their students' queries in the ``Instructor View'' (Figure~\ref{fig:instructor_view}).
This view provides a list of the users in their class with query counts (total and within the past week) and a list of all the student queries.
The list of queries shows salient details of each query (with full text for any field appearing when hovering the cursor over it), and any row can be selected to take the instructor to the response view for that query.
The list of queries can be filtered to show those from a selected user, and it is searchable (full text) and sortable.
Instructors can also download their class data as CSV files.

\toolname{} integrates with learning management systems (LMSes) like Moodle or Canvas that support LTI (Learning Tools Interoperability).
With a small amount of setup, an instructor can provide their students access to \toolname{} via a simple link in their course on the LMS.
Via this link, students may access \toolname{} and be automatically authenticated without having to create, manage, or use a separate login.
Instructors and TAs are identified automatically by LTI, so they have access to the instructor interfaces in \toolname{} with no additional work.
They can then configure their course for student use and monitor their students' queries and the responses they are receiving.

\section{Limitations and Risks}
\label{sec:limitations}
\toolname{} is subject to many of the known limitations and risks of using LLMs.
In particular, completions can be factually incorrect and can include harmful biases.
The problem of inaccuracies in the LLM responses (sometimes called ``hallucination'' or ``confabulation'') is present in \toolname{} with the models it is currently using.
Sometimes, the response contains one or more false statements, and this may confuse or mislead the user.
Users are sensitised to this issue via the prominent notice above each response saying ``Remember: It will not always be correct!''
In our experience, when inaccuracies did occur, they were often in a particular detail of the response, which still gave correct high-level guidance or pointed the user in the right direction.
In our and our students' experiences, the rate of inaccuracies is low enough for the tool to still be valuable and worth the students' time, and as models improve, the accuracy will improve.

LLMs can learn harmful biases such as gender or racial stereotypes from their training data, which can then be reflected in the completions they generate.
This is a well-known and heavily studied issue in language model research~\cite{weidinger2022taxonomy}, and it has been an important issue to the computing education community as well~\cite{becker2023programming}.
While the models used by \toolname{} have been specifically trained and improved by OpenAI to reduce these biases, some still exist~\cite{zhuo2023red}.
These models generally do not make offensive statements unless one actively crafts a prompt to elicit one, but for example they might respond in a way that implicitly reflects a common stereotype.
This is highly unlikely to occur in the context of requesting help on a specific programming issue, but the possibility exists.

The above issues apply to most LLM-based tools, and the likelihood of an LLM's response being incorrect, harmful, off-topic, or otherwise ``off the rails'' increases with additional rounds of user input and model response.
Therefore, by design, every query to \toolname{} is a one-shot request, independent of any others and with no possibility for follow-up or dialogue.
This limits the usefulness of the system, as asking a follow-up question or requesting additional information in the context of an initial response could be very helpful, but the one-shot limitation is imposed to mitigate many of the risks of using LLMs.
Users can submit revised queries with additional information or questions informed by an earlier response if they choose to.

\section{Experiences and Results}
\label{sec:experience}

We used \toolname{} in two sections of an undergraduate introductory-level computer- and data-science course taught by an author of this paper in the Spring semester of 2023. Fifty two students completed the course. Of those students, our analyses includes data from 49 who used \toolname{} at least once during the semester, and data from 45 who completed a survey about using \toolname{} at the end of the semester.
The course is designed to serve a broad audience and attracts students from across the institution who take the course to meet general education requirements or to meet requirements for data-analytic or data-science related credentials. 

The course provides twelve weeks of instruction in Python foundations and three weeks of instruction in Pandas\footnote{Pandas. Available at: \url{https://pandas.pydata.org/} [accessed 2023-06-20]} and Seaborn\footnote{Seaborn. Available at: \url{https://seaborn.pydata.org/} [accessed 2023-06-20]}.
The format of the course is ``flipped,'' with students responsible for reading course materials prior to class, while class time is spent working through assignments on lab computers.
The instructor and a TA assist students and provide instruction/support as needed. \toolname{} was introduced in the fourth week of the semester with a quick demonstration in class. During class, students were encouraged to use \toolname{} for assistance first before asking the instructor or TA for help, but they were otherwise free to make their own choices about when and how to use it. 

\subsection{Student Use}

\begin{figure*}
    \centering
    \begin{minipage}{.47\linewidth}
        \includegraphics[width=\linewidth]{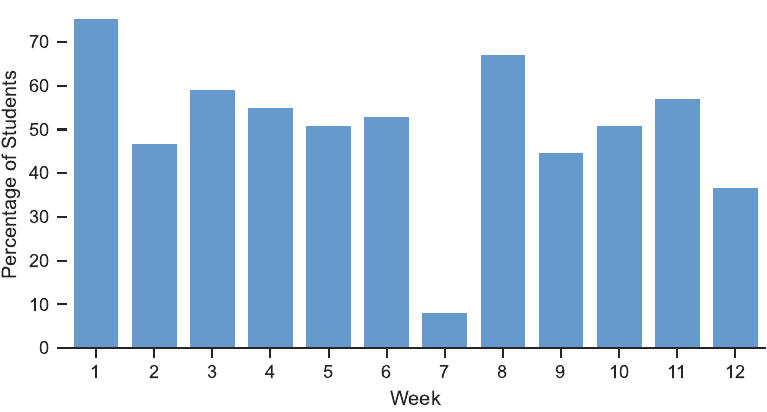}
        \caption{Percentage of the class (y axis) using \toolname{} each week (x axis) across the semester [7 = spring break]. Note that the y axis scale only extends to 70. The figure shows consistent use across the whole semester.}
        \label{fig:usage_by_week}
    \end{minipage}
    \hspace{.05\linewidth}
    \begin{minipage}{.47\linewidth}
        \includegraphics[width=\linewidth]{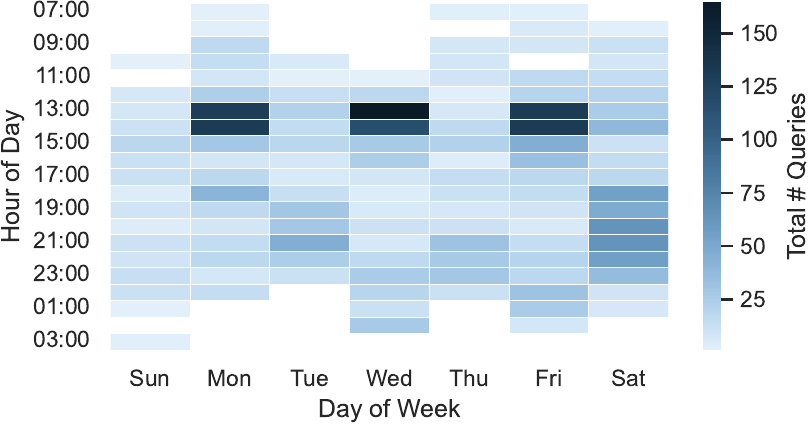}
        \caption{Queries by hour (y axis) and day (x axis) over the whole term. The time span between 4 and 7 AM is not shown due to no activity. The high activity blocks on Mon, Wed, and Fri correspond to the times students were in the classroom. The higher activity on Saturday evening is prior to a recurring deadline for weekly assignments.}
        \label{fig:usage_heatmap}
    \end{minipage}
\end{figure*}

Even with no firm requirement to do so, students used \toolname{} consistently throughout the semester.
Figure~\ref{fig:usage_by_week} shows that roughly half of the class used \toolname{} each week, and we saw that roughly 70\% of the students used \toolname{} in four or more different weeks.
We also observed a wide range of intensity of use between students.
Roughly 80\% of the class submitted 10 or more queries (indicating more than initial trial usage), roughly 50\% submitted 30 or more, and seven of the 49 submitted over 100 queries, including one student with more than 600 queries.
The heatmap in Figure~\ref{fig:usage_heatmap} shows the usage concentrated during two separate class sessions (1 and 2pm on Mon/Wed/Fri) and before assignments were due on Saturday.
Otherwise, there was some use across nearly all hours, including many when no instructor or TA would have been available.
Overall, the continuing, consistent usage strongly suggests that the students generally found the tool beneficial.

\subsection{Student Survey}
At the end of the course we distributed an online survey to understand students' perceptions of \toolname{}. Taking the survey was optional, but students did receive extra-credit for completing it. A total of 45 students (87 percent of the class) completed the survey.
Table~\ref{tbl:survey_results} shows the results for a selection of questions about students' perceptions of the tool and its value to them.
Overall, students found it valuable, and a large majority (95\%) were interested in using it in future CS courses.

\begin{table*}
    \caption{Results for selected questions in the student survey ($n=45$ of 52 students). Rows may not sum to 100\% due to rounding.}
    \label{tbl:survey_results}
    \newcolumntype{L}{>{\centering\arraybackslash}m{1.2cm}}
    \begin{tabular}{p{4.5cm}LLLL}
        \toprule
        & Strongly Agree & Agree & Disagree & Strongly Disagree \\
        \midrule
        \toolname{} helped me complete my work successfully. & 9\% & 71\% & 18\% & 2\% \\
        \toolname{} helped me learn the course material. & 7\% & 56\% & 33\% & 4\% \\
        If I took more Computer Science courses, I would like to be able to use \toolname{} in those classes. & 31\% & 64\% & 4\% & 0\% \\
        \bottomrule
    \end{tabular}
\end{table*}

For additional detail, the survey included the following open-response questions, which were designed to elicit both positive and negative responses:

\begin{itemize}
    \item Q1: What did you find most beneficial about using \toolname{}?
    \item Q2: Do you think there is anything negative about students using \toolname{}?
\end{itemize}

In general, responses were relatively short but tended to be longer for the first question on beneficial aspects (word count; M = 16.2, SD = 10.3) compared to the second question on negative aspects (M = 12.0, SD = 13.0).  To understand the patterns present in the responses, we conducted a thematic analysis in which interesting features of each response were extracted as codes and then collated into higher-level themes \cite{braun2006using}.  We identified five prominent themes in the response to Q1, highlighted in bold in the text that follows.  

The most prominent theme by a clear margin, appearing in 19 of the student responses, was around {\bf``availability''} and specifically that students valued the convenience of being able to ask for assistance outside of the classroom when TAs and the professor were busy or unavailable.  Responses representative of this theme include: \emph{``it was a tool that was always there when I needed it, I didn't have to go to office or TA hours or email''} and \emph{``the ability to get help without talking to professor or TA''}.

Many students (11) explicitly appreciated that \toolname{} could aid them in {\bf``fixing errors''}, which was the next most common theme.  This included getting help to understand error messages and producing explanations of errors.  The following are two examples of typical quotes supporting this theme:  \emph{``it was helpful in understanding some of the error message we hadn't learned about in class''} and \emph{``it really helps with trouble shooting when it comes to semantic errors''}.

One interesting theme that emerged (10 students), distinct from the ``availability'' of \toolname{}, was that it supported {\bf``independence''} by enabling students to make progress without the need to seek external help when they were stuck.  This included providing initial support to students who had difficulty starting work, nudging students in the right direction when they were close to a solution, and helping students who were anxious to ask for help without the fear of embarrassment.  Comments that supported this theme included \emph{``It was nice to have a source to ask when I was unsure how to begin coding''}, \emph{``it helped lead me in the right direction if I almost had the right code''} and \emph{``I felt like I could ask it any question, even dumb ones, which I often did to avoid embarrassing myself in front of the Professor or TA''}.

The remaining themes, which were less common, focused on the {\bf``speed''} (6) with which students could make progress or obtain feedback and the use of \toolname{} to assist with {\bf``learning/understanding''} (7).  Typical comments aligning with these themes included\emph{``Helped me work faster''} and \emph{``it helped understand the code I was writing sometimes''}.  Students also appreciated that \toolname{} would provide guidance rather than directly revealing the solution, as exemplified by the comment \emph{``It gave us help on the answer not just the answer itself''}.   Overall, the responses to Q1 tell a story that \toolname{} was seen as a useful resource for obtaining rapid assistance and a complementary tool to traditional TA and instructor support.

As to the concerns (Q2), we also identified five prominent themes, again highlighted in bold. Around half of the students (24) stated that they had \textbf{``no concerns''}. Some of the students would even suggest the use of the tool should have been more extensive: \emph{``We should even use it during quizzes''}. Others explained why they did not have any concerns: \emph{``No, absolutely not, especially considering it never handed me the answer on a silver platter.''}

The most prominent theme as to the concerns was the perceived \textbf{``difficulty''} in using \toolname{}. Multiple students (14) stated that the tool is difficult to use when the problem is not understood: \emph{``sometimes i didnt know exactly what to ask.. but i usually got there eventually''} and \emph{``I did not like how hard it was to ask something I do not understand.''}. Several students also reported receiving answers that were difficult to utilize or not helpful: \emph{``There were many times that \toolname{} misunderstood my question and gave me advice which confused me even more.''} and \emph{``Sometimes it gives really strange responses that are not related to the problem''}.

Several students (5) reported that sometimes an answer provided by \toolname{} contained elements that were \textbf{``not covered''} in class and, hence, the students were not expected to have knowledge of those elements. Responses representative of this theme included: \emph{``Sometimes it tells you to do code that we haven't learned in class''} and \emph{``I would run into the issue where it wanted me to use concepts that I haven't been taught yet. This is both and good and a bad thing because it can introduce students to resources, but also confuse them.''}.

A small number of students' responses (3) were hinting on using \toolname{} without investing proper effort at solving the problem independently (i.e., \textbf{``over-reliance''}). The responses suggest that the students were aware this could have negative effects on their learning, yet, they would still engage in that practice: \emph{`` think some people could complete the code without help and by going directly to \toolname{} their limiting themselves''} and \emph{``I do think that sometimes I can get to dependent on \toolname{} and I have to scale it back a bit.''}.

Several responses (3) stated that \toolname{} is \textbf{``not human''} and, hence, its capabilities are in some way limited as compared to the assistance provided by an instructor or a TA. However, the responses do not go into much detail as why this might be the case: \emph{``less personal''} and \emph{``No, but it cannot be a substitute for a real person.''} One of the responses explained the preference for human assistance in terms of difficulty (see above) of formulating the proper question for \toolname{}: \emph{``no but personally I prefer to ask a real person because its difficult to phrase you questions in a way that won't confuse \toolname{}''}.

\subsection{Instructor Reflections}
After the conclusion of the semester, the instructor, who is also one of the authors, reflected on what did and did not work:

\toolname{} was easy to introduce to the class. As an instructional resource, its utility is immediately and obviously apparent. Students required little convincing to give it a try. While in class, we requested that students ask \toolname{} for help before seeking help from the instructor or teaching assistant. We did not enforce this as a rule but encouraged it throughout the semester. The idea was that \toolname{} could provide an initial level of support and handle relatively straightforward but common concerns, such as syntax errors. \toolname{} performed very well in this capacity, and given its flexibility and low-cost, it is a great addition to the classroom for this functionality alone. However, \toolname{} also provided much more sophisticated help on a huge range of introductory CS problems throughout the semester.

\toolname{} appeared to provide accurate and helpful responses to students the majority of the time. \toolname{} did not “give away the answer” or otherwise become a complete replacement for actively working through problems. It appears to strike a nice balance between providing enough information to move students forward without undermining the intent of the assignments.

\toolname{} was a great addition to the course in terms of serving students who had difficulty attending office hours or who needed frequent reassurance or feedback as they worked through assignments outside of class time. It was also exceptional in providing a novel avenue for delivering support to students who did not take advantage of traditional avenues of support. For example, some students who seemed uncomfortable, embarrassed, or otherwise reluctant to ask for help from the instructor or TA had no reservations about asking \toolname{}. 

\toolname{} sometimes provided assistance that was inconsistent with the content of the class and the knowledge-level of the students. For example, \toolname{} might suggest solving problems with methods that had not yet been introduced. This was confusing and frustrating for some students. During the semester, the avoid set functionality (Section~\ref{sec:instructors}) was added to allow the instructor to explicitly prohibit certain kinds of content in \toolname{} responses, which largely resolved the problem. Students sometimes provided too little information describing their problem to get a useful response and required some coaching to provide detailed or thoughtful descriptions of problems to \toolname{}. 

Reviewing student queries submitted to \toolname{} provided an entirely new type of insight into student learning. In comparison to submitted work, the queries were a much more direct and unfiltered look into student thinking as they worked through problems. On some occasions, this feedback guided modifications of assignments and additional class instruction during the semester. 

Overall, given its great utility in a wide range of circumstances, its ease of use, and low cost, I found \toolname{} to be a tremendous asset in my course. I intend to continue using it in all of my introductory courses moving forward.

\section{Recommended Practices}
\label{sec:recommended_practices}

Based on our experiences, we have collected a few recommendations for integrating \toolname{} into a class effectively.

\paragraph{Initial introduction}
When first introducing \toolname{} to students, motivate its use by sharing some of the benefits identified in this work, as relevant to your course.
Explain carefully its strengths and limitations in the context of your course: how it will likely be able to help, and where may it produce incorrect responses.
Provide guidance on how to ask for help most \emph{effectively}.
This includes providing the relevant portions of one's code, identifying and copying the important information from error messages, and providing enough information for the issue to be identified.
These are the same skills one needs to effectively communicate issues to instructors or peers.
Providing good and bad examples or taking a moment to roleplay a few situations may help here.
Demonstrate \toolname{} with a few issues similar to those you expect your students to encounter.
Model how to provide sufficient information and communicate clearly.

\paragraph{During Use}
Throughout the course, while students are using \toolname{}, it is helpful to view the students' queries regularly.
You can gain detailed insight into where they are struggling at each point in the term that may lead to adapting course plans.
Additionally, you might identify students whose usage is not effective (e.g., repeatedly submitting ineffective queries or demonstrating over-reliance), and reach out to them directly to provide guidance or a nudge.

Instructors and TAs should sample \toolname{}'s responses in each section of the course to spot and mitigate issues.
For example, if \toolname{} suggests a technique, function, or concept that does not fit the design of your course, you can add that to the avoid set (Section~\ref{sec:instructors}) to prevent it from being used in future responses.

\section{Conclusion and Future Work}
\label{sec:conclusion}

This work shows that LLMs, when properly implemented and integrated into a learning environment, can be a valuable aid to both students and educators.
We developed \toolname{} to provide immediate, high-quality support to students working on programming exercises while mitigating the risk of fostering an over-reliance on the automated assistance.
Providing an automated option for this kind of help can increase the level of support students receive throughout a course due to a combination of being constantly available and avoiding the anxiety associated with asking a professor or TA for help.
In our pilot study, students found \toolname{} to be a welcome addition to direct support from a professor and teaching assistants.

Going forward, we intend to continue developing and improving \toolname{}.
The ``avoid set'' functionality proved to be critical for obtaining course-appropriate responses in many cases, and we plan to give instructors more ways to provide context about their courses and thus further tailor the LLM responses for their students.
Additionally, we plan to explore different forms or levels of intervention that might be appropriate depending on the complexity of the task, the experience level of the student, or even the specific learning objectives of the course.
And we see many opportunities for the tool to be more individualized, adapting to the needs of each student.
For example, it could record and maintain information about each individual student's mastery of different topics, using that to guide the responses generated for them.

While encouraging, this work presents only an initial exploration into the effective deployment of LLMs in computing education. For example, while students positively rated \toolname{} and the instructor found it easy to use and deploy, future work should establish more robust metrics for gauging efficacy, such as measuring impact on student learning outcomes or comparing student performance in classrooms that use \toolname{} to those that do not.

We also recognize that further work needs to be conducted with larger, more diverse populations of students. It would also be interesting to deploy \toolname{} in different educational settings, such as in distance learning or self-paced programming courses, to evaluate its flexibility and adaptability.

Our findings could have implications beyond computing education. LLMs such as those used in \toolname{} could potentially be adapted to support learning in other domains. We hope that our work serves as an impetus for other researchers and educators to explore the use of LLMs in diverse educational contexts, continuing the dialogue around the opportunities and challenges they present.

\balance
\bibliographystyle{ACM-Reference-Format}
\bibliography{main}


\begin{thebibliography}{37}


\ifx \showCODEN    \undefined \def \showCODEN     #1{\unskip}     \fi
\ifx \showDOI      \undefined \def \showDOI       #1{#1}\fi
\ifx \showISBNx    \undefined \def \showISBNx     #1{\unskip}     \fi
\ifx \showISBNxiii \undefined \def \showISBNxiii  #1{\unskip}     \fi
\ifx \showISSN     \undefined \def \showISSN      #1{\unskip}     \fi
\ifx \showLCCN     \undefined \def \showLCCN      #1{\unskip}     \fi
\ifx \shownote     \undefined \def \shownote      #1{#1}          \fi
\ifx \showarticletitle \undefined \def \showarticletitle #1{#1}   \fi
\ifx \showURL      \undefined \def \showURL       {\relax}        \fi
\providecommand\bibfield[2]{#2}
\providecommand\bibinfo[2]{#2}
\providecommand\natexlab[1]{#1}
\providecommand\showeprint[2][]{arXiv:#2}

\bibitem[Becker et~al\mbox{.}(2023)]%
        {becker2023programming}
\bibfield{author}{\bibinfo{person}{Brett~A Becker}, \bibinfo{person}{Paul
  Denny}, \bibinfo{person}{James Finnie-Ansley}, \bibinfo{person}{Andrew
  Luxton-Reilly}, \bibinfo{person}{James Prather}, {and}
  \bibinfo{person}{Eddie~Antonio Santos}.} \bibinfo{year}{2023}\natexlab{}.
\newblock \showarticletitle{Programming Is Hard-Or at Least It Used to Be:
  Educational Opportunities and Challenges of AI Code Generation}. In
  \bibinfo{booktitle}{\emph{Proceedings of the 54th ACM Technical Symposium on
  Computer Science Education V. 1}}. \bibinfo{pages}{500--506}.
\newblock


\bibitem[Braun and Clarke(2006)]%
        {braun2006using}
\bibfield{author}{\bibinfo{person}{Virginia Braun} {and}
  \bibinfo{person}{Victoria Clarke}.} \bibinfo{year}{2006}\natexlab{}.
\newblock \showarticletitle{Using thematic analysis in psychology}.
\newblock \bibinfo{journal}{\emph{Qualitative Research in Psychology}}
  \bibinfo{volume}{3}, \bibinfo{number}{2} (\bibinfo{year}{2006}),
  \bibinfo{pages}{77--101}.
\newblock
\urldef\tempurl%
\url{https://doi.org/10.1191/1478088706qp063oa}
\showDOI{\tempurl}


\bibitem[Brusilovsky et~al\mbox{.}(2023)]%
        {brusilovsky2023future}
\bibfield{author}{\bibinfo{person}{Peter Brusilovsky},
  \bibinfo{person}{Barbara~J Ericson}, \bibinfo{person}{Cay~S Horstmann}, {and}
  \bibinfo{person}{Christian Servin}.} \bibinfo{year}{2023}\natexlab{}.
\newblock \showarticletitle{The Future of Computing Education Materials}.
\newblock  (\bibinfo{year}{2023}).
\newblock


\bibitem[Carreira et~al\mbox{.}(2022)]%
        {carreira_pyo_2022}
\bibfield{author}{\bibinfo{person}{Gustavo Carreira}, \bibinfo{person}{Leonardo
  Silva}, \bibinfo{person}{Antonio~Jose Mendes}, {and}
  \bibinfo{person}{Hugo~Goncalo Oliveira}.} \bibinfo{year}{2022}\natexlab{}.
\newblock \showarticletitle{Pyo, a {Chatbot} {Assistant} for {Introductory}
  {Programming} {Students}}. In \bibinfo{booktitle}{\emph{2022 {International}
  {Symposium} on {Computers} in {Education} ({SIIE})}}.
  \bibinfo{publisher}{IEEE}, \bibinfo{address}{Coimbra, Portugal},
  \bibinfo{pages}{1--6}.
\newblock
\showISBNx{978-1-66547-464-1}
\urldef\tempurl%
\url{https://doi.org/10.1109/SIIE56031.2022.9982349}
\showDOI{\tempurl}


\bibitem[Chen et~al\mbox{.}(2022)]%
        {chen_codet_2022}
\bibfield{author}{\bibinfo{person}{Bei Chen}, \bibinfo{person}{Fengji Zhang},
  \bibinfo{person}{Anh Nguyen}, \bibinfo{person}{Daoguang Zan},
  \bibinfo{person}{Zeqi Lin}, \bibinfo{person}{Jian-Guang Lou}, {and}
  \bibinfo{person}{Weizhu Chen}.} \bibinfo{year}{2022}\natexlab{}.
\newblock \bibinfo{title}{{CodeT}: {Code} {Generation} with {Generated}
  {Tests}}.
\newblock
\newblock
\showeprint[arxiv]{2207.10397}~[cs.CL]


\bibitem[Chen et~al\mbox{.}(2021)]%
        {chen2021evaluating}
\bibfield{author}{\bibinfo{person}{Mark Chen}, \bibinfo{person}{Jerry Tworek},
  \bibinfo{person}{Heewoo Jun}, \bibinfo{person}{Qiming Yuan},
  \bibinfo{person}{Henrique Ponde de~Oliveira Pinto}, \bibinfo{person}{Jared
  Kaplan}, \bibinfo{person}{Harri Edwards}, \bibinfo{person}{Yuri Burda},
  \bibinfo{person}{Nicholas Joseph}, \bibinfo{person}{Greg Brockman},
  {et~al\mbox{.}}} \bibinfo{year}{2021}\natexlab{}.
\newblock \bibinfo{title}{Evaluating large language models trained on code}.
\newblock
\newblock
\showeprint[arxiv]{2107.03374}~[cs.LG]


\bibitem[Collins(2023)]%
        {collins2023policy}
\bibfield{author}{\bibinfo{person}{Jonathan~E Collins}.}
  \bibinfo{year}{2023}\natexlab{}.
\newblock \showarticletitle{Policy Solutions: Policy questions for ChatGPT and
  artificial intelligence}.
\newblock \bibinfo{journal}{\emph{Phi Delta Kappan}} \bibinfo{volume}{104},
  \bibinfo{number}{7} (\bibinfo{year}{2023}), \bibinfo{pages}{60--61}.
\newblock


\bibitem[Crow et~al\mbox{.}(2018)]%
        {crow_intelligent_2018}
\bibfield{author}{\bibinfo{person}{Tyne Crow}, \bibinfo{person}{Andrew
  Luxton-Reilly}, {and} \bibinfo{person}{Burkhard Wuensche}.}
  \bibinfo{year}{2018}\natexlab{}.
\newblock \showarticletitle{Intelligent tutoring systems for programming
  education: a systematic review}. In \bibinfo{booktitle}{\emph{Proceedings of
  the 20th {Australasian} {Computing} {Education} {Conference}}}.
  \bibinfo{publisher}{ACM}, \bibinfo{address}{Brisbane Queensland Australia},
  \bibinfo{pages}{53--62}.
\newblock
\showISBNx{978-1-4503-6340-2}
\urldef\tempurl%
\url{https://doi.org/10.1145/3160489.3160492}
\showDOI{\tempurl}


\bibitem[Denny et~al\mbox{.}(2023a)]%
        {denny_conversing_2023}
\bibfield{author}{\bibinfo{person}{Paul Denny}, \bibinfo{person}{Viraj Kumar},
  {and} \bibinfo{person}{Nasser Giacaman}.} \bibinfo{year}{2023}\natexlab{a}.
\newblock \showarticletitle{Conversing with {Copilot}: {Exploring} {Prompt}
  {Engineering} for {Solving} {CS1} {Problems} {Using} {Natural} {Language}}.
  In \bibinfo{booktitle}{\emph{Proceedings of the 54th {ACM} {Technical}
  {Symposium} on {Computer} {Science} {Education} {V}. 1}}.
  \bibinfo{publisher}{ACM}, \bibinfo{address}{Toronto ON Canada},
  \bibinfo{pages}{1136--1142}.
\newblock
\showISBNx{978-1-4503-9431-4}
\urldef\tempurl%
\url{https://doi.org/10.1145/3545945.3569823}
\showDOI{\tempurl}


\bibitem[Denny et~al\mbox{.}(2023b)]%
        {denny2023promptly}
\bibfield{author}{\bibinfo{person}{Paul Denny}, \bibinfo{person}{Juho
  Leinonen}, \bibinfo{person}{James Prather}, \bibinfo{person}{Andrew
  Luxton-Reilly}, \bibinfo{person}{Thezyrie Amarouche},
  \bibinfo{person}{Brett~A. Becker}, {and} \bibinfo{person}{Brent~N. Reeves}.}
  \bibinfo{year}{2023}\natexlab{b}.
\newblock \bibinfo{title}{Promptly: Using Prompt Problems to Teach Learners How
  to Effectively Utilize AI Code Generators}.
\newblock
\newblock
\showeprint[arxiv]{2307.16364}~[cs.HC]


\bibitem[Denny et~al\mbox{.}(2023c)]%
        {denny2023computing}
\bibfield{author}{\bibinfo{person}{Paul Denny}, \bibinfo{person}{James
  Prather}, \bibinfo{person}{Brett~A. Becker}, \bibinfo{person}{James
  Finnie-Ansley}, \bibinfo{person}{Arto Hellas}, \bibinfo{person}{Juho
  Leinonen}, \bibinfo{person}{Andrew Luxton-Reilly}, \bibinfo{person}{Brent~N.
  Reeves}, \bibinfo{person}{Eddie~Antonio Santos}, {and} \bibinfo{person}{Sami
  Sarsa}.} \bibinfo{year}{2023}\natexlab{c}.
\newblock \bibinfo{title}{Computing Education in the Era of Generative AI}.
\newblock
\newblock
\showeprint[arxiv]{2306.02608}~[cs.CY]


\bibitem[Finnie-Ansley et~al\mbox{.}(2022)]%
        {finnie2022robots}
\bibfield{author}{\bibinfo{person}{James Finnie-Ansley}, \bibinfo{person}{Paul
  Denny}, \bibinfo{person}{Brett~A Becker}, \bibinfo{person}{Andrew
  Luxton-Reilly}, {and} \bibinfo{person}{James Prather}.}
  \bibinfo{year}{2022}\natexlab{}.
\newblock \showarticletitle{The robots are coming: Exploring the implications
  of openai codex on introductory programming}. In
  \bibinfo{booktitle}{\emph{Proceedings of the 24th Australasian Computing
  Education Conference}}. \bibinfo{pages}{10–19}.
\newblock
\urldef\tempurl%
\url{https://doi.org/10.1145/3511861.3511863}
\showDOI{\tempurl}


\bibitem[Gao et~al\mbox{.}(2022)]%
        {gao2022who}
\bibfield{author}{\bibinfo{person}{Zhikai Gao}, \bibinfo{person}{Sarah
  Heckman}, {and} \bibinfo{person}{Collin Lynch}.}
  \bibinfo{year}{2022}\natexlab{}.
\newblock \showarticletitle{Who Uses Office Hours? A Comparison of In-Person
  and Virtual Office Hours Utilization}. In
  \bibinfo{booktitle}{\emph{Proceedings of the 53rd ACM Technical Symposium on
  Computer Science Education - Volume 1}} (Providence, RI, USA)
  \emph{(\bibinfo{series}{SIGCSE 2022})}. \bibinfo{publisher}{Association for
  Computing Machinery}, \bibinfo{address}{New York, NY, USA},
  \bibinfo{pages}{300–306}.
\newblock
\showISBNx{9781450390705}
\urldef\tempurl%
\url{https://doi.org/10.1145/3478431.3499334}
\showDOI{\tempurl}


\bibitem[Hellas et~al\mbox{.}(2023)]%
        {hellas2023exploring}
\bibfield{author}{\bibinfo{person}{Arto Hellas}, \bibinfo{person}{Juho
  Leinonen}, \bibinfo{person}{Sami Sarsa}, \bibinfo{person}{Charles Koutcheme},
  \bibinfo{person}{Lilja Kujanp{\"a}{\"a}}, {and} \bibinfo{person}{Juha
  Sorva}.} \bibinfo{year}{2023}\natexlab{}.
\newblock \bibinfo{title}{Exploring the Responses of Large Language Models to
  Beginner Programmers' Help Requests}.
\newblock
\newblock
\showeprint[arxiv]{2306.05715}~[cs.CY]


\bibitem[Jalil et~al\mbox{.}(2023)]%
        {jalil_chatgpt_2023}
\bibfield{author}{\bibinfo{person}{Sajed Jalil}, \bibinfo{person}{Suzzana
  Rafi}, \bibinfo{person}{Thomas~D. LaToza}, \bibinfo{person}{Kevin Moran},
  {and} \bibinfo{person}{Wing Lam}.} \bibinfo{year}{2023}\natexlab{}.
\newblock \showarticletitle{{ChatGPT} and Software Testing Education: Promises
  {\&} Perils}. In \bibinfo{booktitle}{\emph{2023 {IEEE} International
  Conference on Software Testing, Verification and Validation Workshops
  ({ICSTW})}}. \bibinfo{publisher}{{IEEE}}.
\newblock
\urldef\tempurl%
\url{https://doi.org/10.1109/icstw58534.2023.00078}
\showDOI{\tempurl}
\showeprint{arXiv:2302.03287}


\bibitem[Kasneci et~al\mbox{.}(2023)]%
        {kasneci2023chatgpt}
\bibfield{author}{\bibinfo{person}{Enkelejda Kasneci}, \bibinfo{person}{Kathrin
  Sessler}, \bibinfo{person}{Stefan Küchemann}, \bibinfo{person}{Maria
  Bannert}, \bibinfo{person}{Daryna Dementieva}, \bibinfo{person}{Frank
  Fischer}, \bibinfo{person}{Urs Gasser}, \bibinfo{person}{Georg Groh},
  \bibinfo{person}{Stephan Günnemann}, \bibinfo{person}{Eyke Hüllermeier},
  \bibinfo{person}{Stepha Krusche}, \bibinfo{person}{Gitta Kutyniok},
  \bibinfo{person}{Tilman Michaeli}, \bibinfo{person}{Claudia Nerdel},
  \bibinfo{person}{Jürgen Pfeffer}, \bibinfo{person}{Oleksandra Poquet},
  \bibinfo{person}{Michael Sailer}, \bibinfo{person}{Albrecht Schmidt},
  \bibinfo{person}{Tina Seidel}, \bibinfo{person}{Matthias Stadler},
  \bibinfo{person}{Jochen Weller}, \bibinfo{person}{Jochen Kuhn}, {and}
  \bibinfo{person}{Gjergji Kasneci}.} \bibinfo{year}{2023}\natexlab{}.
\newblock \showarticletitle{ChatGPT for good? On opportunities and challenges
  of large language models for education}.
\newblock \bibinfo{journal}{\emph{Learning and Individual Differences}}
  \bibinfo{volume}{103} (\bibinfo{year}{2023}), \bibinfo{pages}{102274}.
\newblock
\showISSN{1041-6080}
\urldef\tempurl%
\url{https://doi.org/10.1016/j.lindif.2023.102274}
\showDOI{\tempurl}


\bibitem[Kazemitabaar et~al\mbox{.}(2023)]%
        {kazemitabaar2023studying}
\bibfield{author}{\bibinfo{person}{Majeed Kazemitabaar},
  \bibinfo{person}{Justin Chow}, \bibinfo{person}{Carl Ka~To Ma},
  \bibinfo{person}{Barbara~J. Ericson}, \bibinfo{person}{David Weintrop}, {and}
  \bibinfo{person}{Tovi Grossman}.} \bibinfo{year}{2023}\natexlab{}.
\newblock \showarticletitle{Studying the Effect of AI Code Generators on
  Supporting Novice Learners in Introductory Programming}. In
  \bibinfo{booktitle}{\emph{Proceedings of the 2023 CHI Conference on Human
  Factors in Computing Systems}} (Hamburg, Germany) \emph{(\bibinfo{series}{CHI
  '23})}. \bibinfo{publisher}{Association for Computing Machinery},
  \bibinfo{address}{New York, NY, USA}, Article \bibinfo{articleno}{455},
  \bibinfo{numpages}{23}~pages.
\newblock
\showISBNx{9781450394215}
\urldef\tempurl%
\url{https://doi.org/10.1145/3544548.3580919}
\showDOI{\tempurl}


\bibitem[Keuning et~al\mbox{.}(2019)]%
        {keuning_systematic_2019}
\bibfield{author}{\bibinfo{person}{Hieke Keuning}, \bibinfo{person}{Johan
  Jeuring}, {and} \bibinfo{person}{Bastiaan Heeren}.}
  \bibinfo{year}{2019}\natexlab{}.
\newblock \showarticletitle{A {Systematic} {Literature} {Review} of {Automated}
  {Feedback} {Generation} for {Programming} {Exercises}}.
\newblock \bibinfo{journal}{\emph{ACM Transactions on Computing Education}}
  \bibinfo{volume}{19}, \bibinfo{number}{1} (\bibinfo{date}{March}
  \bibinfo{year}{2019}), \bibinfo{pages}{1--43}.
\newblock
\showISSN{1946-6226}
\urldef\tempurl%
\url{https://doi.org/10.1145/3231711}
\showDOI{\tempurl}


\bibitem[Konecki et~al\mbox{.}(2015)]%
        {konecki_intelligent_2015}
\bibfield{author}{\bibinfo{person}{Mario Konecki}, \bibinfo{person}{Nikola
  Kadoic}, {and} \bibinfo{person}{Rok Piltaver}.}
  \bibinfo{year}{2015}\natexlab{}.
\newblock \showarticletitle{Intelligent assistant for helping students to learn
  programming}. In \bibinfo{booktitle}{\emph{2015 38th {International}
  {Convention} on {Information} and {Communication} {Technology}, {Electronics}
  and {Microelectronics} ({MIPRO})}}. \bibinfo{publisher}{IEEE},
  \bibinfo{address}{Opatija, Croatia}, \bibinfo{pages}{924--928}.
\newblock
\showISBNx{978-953-233-082-3}
\urldef\tempurl%
\url{https://doi.org/10.1109/MIPRO.2015.7160406}
\showDOI{\tempurl}


\bibitem[Leinonen et~al\mbox{.}(2023)]%
        {leinonen_comparing_2023}
\bibfield{author}{\bibinfo{person}{Juho Leinonen}, \bibinfo{person}{Paul
  Denny}, \bibinfo{person}{Stephen MacNeil}, \bibinfo{person}{Sami Sarsa},
  \bibinfo{person}{Seth Bernstein}, \bibinfo{person}{Joanne Kim},
  \bibinfo{person}{Andrew Tran}, {and} \bibinfo{person}{Arto Hellas}.}
  \bibinfo{year}{2023}\natexlab{}.
\newblock \bibinfo{title}{Comparing {Code} {Explanations} {Created} by
  {Students} and {Large} {Language} {Models}}.
\newblock
\newblock
\showeprint[arxiv]{2304.03938}~[cs.CY]


\bibitem[Mahdaoui et~al\mbox{.}(2022)]%
        {mahdaoui_comparative_2022}
\bibfield{author}{\bibinfo{person}{Mariam Mahdaoui}, \bibinfo{person}{Said
  Nouh}, \bibinfo{person}{My~Seddiq~ELKASMI Alaoui}, {and}
  \bibinfo{person}{Mounir Sadiq}.} \bibinfo{year}{2022}\natexlab{}.
\newblock \showarticletitle{Comparative study between automatic hint generation
  approaches in {Intelligent} {Programming} {Tutors}}.
\newblock \bibinfo{journal}{\emph{Procedia Computer Science}}
  \bibinfo{volume}{198} (\bibinfo{year}{2022}), \bibinfo{pages}{391--396}.
\newblock
\showISSN{18770509}
\urldef\tempurl%
\url{https://doi.org/10.1016/j.procs.2021.12.259}
\showDOI{\tempurl}


\bibitem[McBroom et~al\mbox{.}(2022)]%
        {mcbroom_survey_2022}
\bibfield{author}{\bibinfo{person}{Jessica McBroom}, \bibinfo{person}{Irena
  Koprinska}, {and} \bibinfo{person}{Kalina Yacef}.}
  \bibinfo{year}{2022}\natexlab{}.
\newblock \showarticletitle{A {Survey} of {Automated} {Programming} {Hint}
  {Generation}: {The} {HINTS} {Framework}}.
\newblock \bibinfo{journal}{\emph{Comput. Surveys}} \bibinfo{volume}{54},
  \bibinfo{number}{8} (\bibinfo{date}{Nov.} \bibinfo{year}{2022}),
  \bibinfo{pages}{1--27}.
\newblock
\showISSN{0360-0300, 1557-7341}
\urldef\tempurl%
\url{https://doi.org/10.1145/3469885}
\showDOI{\tempurl}


\bibitem[Nguyen and Nadi(2022)]%
        {nguyen_empirical_2022}
\bibfield{author}{\bibinfo{person}{Nhan Nguyen} {and} \bibinfo{person}{Sarah
  Nadi}.} \bibinfo{year}{2022}\natexlab{}.
\newblock \showarticletitle{An empirical evaluation of {GitHub} copilot's code
  suggestions}. In \bibinfo{booktitle}{\emph{Proceedings of the 19th
  {International} {Conference} on {Mining} {Software} {Repositories}}}.
  \bibinfo{publisher}{ACM}, \bibinfo{address}{Pittsburgh Pennsylvania},
  \bibinfo{pages}{1--5}.
\newblock
\showISBNx{978-1-4503-9303-4}
\urldef\tempurl%
\url{https://doi.org/10.1145/3524842.3528470}
\showDOI{\tempurl}


\bibitem[Okonkwo and Ade-Ibijola(2021)]%
        {okonkwo_python-bot_2021}
\bibfield{author}{\bibinfo{person}{Chinedu~Wilfred Okonkwo} {and}
  \bibinfo{person}{Abejide Ade-Ibijola}.} \bibinfo{year}{2021}\natexlab{}.
\newblock \showarticletitle{Python-{Bot}: {A} {Chatbot} for {Teaching} {Python}
  {Programming}}.
\newblock \bibinfo{journal}{\emph{Engineering Letters}}  \bibinfo{volume}{29}
  (\bibinfo{date}{02} \bibinfo{year}{2021}), \bibinfo{pages}{25--34}.
\newblock


\bibitem[Okonkwo and Ade-Ibijola(2022)]%
        {okonkwo_revision-bot_2022}
\bibfield{author}{\bibinfo{person}{Chinedu~Wilfred Okonkwo} {and}
  \bibinfo{person}{Abejide Ade-Ibijola}.} \bibinfo{year}{2022}\natexlab{}.
\newblock \showarticletitle{Revision-{Bot}: {A} {Chatbot} for {Studying} {Past}
  {Questions} in {Introductory} {Programming}}.
\newblock \bibinfo{journal}{\emph{IAENG International Journal of Computer
  Science}} \bibinfo{volume}{49}, \bibinfo{number}{3} (\bibinfo{year}{2022}).
\newblock


\bibitem[Pardos and Bhandari(2023)]%
        {pardos2023learning}
\bibfield{author}{\bibinfo{person}{Zachary~A. Pardos} {and}
  \bibinfo{person}{Shreya Bhandari}.} \bibinfo{year}{2023}\natexlab{}.
\newblock \bibinfo{title}{Learning gain differences between ChatGPT and human
  tutor generated algebra hints}.
\newblock
\newblock
\showeprint[arxiv]{2302.06871}~[cs.CY]


\bibitem[Prather et~al\mbox{.}(2023a)]%
        {prather2023transformed}
\bibfield{author}{\bibinfo{person}{James Prather}, \bibinfo{person}{Paul
  Denny}, \bibinfo{person}{Juho Leinonen}, \bibinfo{person}{Brett~A Becker},
  \bibinfo{person}{Ibrahim Albluwi}, \bibinfo{person}{Michael~E Caspersen},
  \bibinfo{person}{Michelle Craig}, \bibinfo{person}{Hieke Keuning},
  \bibinfo{person}{Natalie Kiesler}, \bibinfo{person}{Tobias Kohn},
  {et~al\mbox{.}}} \bibinfo{year}{2023}\natexlab{a}.
\newblock \showarticletitle{Transformed by Transformers: Navigating the AI
  Coding Revolution for Computing Education: An ITiCSE Working Group Conducted
  by Humans}. In \bibinfo{booktitle}{\emph{Proceedings of the 2023 Conference
  on Innovation and Technology in Computer Science Education V. 2}}.
  \bibinfo{pages}{561--562}.
\newblock


\bibitem[Prather et~al\mbox{.}(2023b)]%
        {prather_its_2023}
\bibfield{author}{\bibinfo{person}{James Prather}, \bibinfo{person}{Brent~N.
  Reeves}, \bibinfo{person}{Paul Denny}, \bibinfo{person}{Brett~A. Becker},
  \bibinfo{person}{Juho Leinonen}, \bibinfo{person}{Andrew Luxton-Reilly},
  \bibinfo{person}{Garrett Powell}, \bibinfo{person}{James Finnie-Ansley},
  {and} \bibinfo{person}{Eddie~Antonio Santos}.}
  \bibinfo{year}{2023}\natexlab{b}.
\newblock \bibinfo{title}{"{It}'s {Weird} {That} it {Knows} {What} {I} {Want}":
  {Usability} and {Interactions} with {Copilot} for {Novice} {Programmers}}.
\newblock
\newblock
\showeprint[arxiv]{2304.02491}~[cs.HC]


\bibitem[Rutgers(2021)]%
        {rutgers_duckbot_nodate}
\bibfield{author}{\bibinfo{person}{Margot Rutgers}.}
  \bibinfo{year}{2021}\natexlab{}.
\newblock \emph{\bibinfo{title}{Duckbot: {A} chatbot to assist students in
  programming tutorials}}.
\newblock \bibinfo{thesistype}{Master's\ thesis}. \bibinfo{school}{University
  of Twente}.
\newblock


\bibitem[Sarsa et~al\mbox{.}(2022)]%
        {sarsa_automatic_2022}
\bibfield{author}{\bibinfo{person}{Sami Sarsa}, \bibinfo{person}{Paul Denny},
  \bibinfo{person}{Arto Hellas}, {and} \bibinfo{person}{Juho Leinonen}.}
  \bibinfo{year}{2022}\natexlab{}.
\newblock \showarticletitle{Automatic {Generation} of {Programming} {Exercises}
  and {Code} {Explanations} {Using} {Large} {Language} {Models}}. In
  \bibinfo{booktitle}{\emph{Proceedings of the 2022 {ACM} {Conference} on
  {International} {Computing} {Education} {Research} {V}.1}}.
  \bibinfo{publisher}{ACM}, \bibinfo{address}{Lugano and Virtual Event
  Switzerland}, \bibinfo{pages}{27--43}.
\newblock
\showISBNx{978-1-4503-9194-8}
\urldef\tempurl%
\url{https://doi.org/10.1145/3501385.3543957}
\showDOI{\tempurl}


\bibitem[Savelka et~al\mbox{.}(2023a)]%
        {savelka2023thrilled}
\bibfield{author}{\bibinfo{person}{Jaromir Savelka}, \bibinfo{person}{Arav
  Agarwal}, \bibinfo{person}{Marshall An}, \bibinfo{person}{Chris Bogart},
  {and} \bibinfo{person}{Majd Sakr}.} \bibinfo{year}{2023}\natexlab{a}.
\newblock \showarticletitle{Thrilled by Your Progress! Large Language Models
  (GPT-4) No Longer Struggle to Pass Assessments in Higher Education
  Programming Course}. In \bibinfo{booktitle}{\emph{Proceedings of the 2023 ACM
  Conference on International Computing Education Research V.1}}.
  \bibinfo{publisher}{ACM}.
\newblock


\bibitem[Savelka et~al\mbox{.}(2023b)]%
        {savelka2023large}
\bibfield{author}{\bibinfo{person}{Jaromir Savelka}, \bibinfo{person}{Arav
  Agarwal}, \bibinfo{person}{Christopher Bogart}, {and} \bibinfo{person}{Majd
  Sakr}.} \bibinfo{year}{2023}\natexlab{b}.
\newblock \bibinfo{title}{Large Language Models (GPT) Struggle to Answer
  Multiple-Choice Questions about Code}.
\newblock
\newblock
\showeprint[arxiv]{2303.08033}~[cs.CL]


\bibitem[Tian et~al\mbox{.}(2023)]%
        {tian_is_2023}
\bibfield{author}{\bibinfo{person}{Haoye Tian}, \bibinfo{person}{Weiqi Lu},
  \bibinfo{person}{Tsz~On Li}, \bibinfo{person}{Xunzhu Tang},
  \bibinfo{person}{Shing-Chi Cheung}, \bibinfo{person}{Jacques Klein}, {and}
  \bibinfo{person}{Tegawendé~F. Bissyandé}.} \bibinfo{year}{2023}\natexlab{}.
\newblock \bibinfo{title}{Is {ChatGPT} the {Ultimate} {Programming} {Assistant}
  -- {How} far is it?}
\newblock
\newblock
\showeprint[arxiv]{2304.11938}~[cs.SE]


\bibitem[Walden et~al\mbox{.}(2022)]%
        {walden_chatbot_2022}
\bibfield{author}{\bibinfo{person}{James Walden}, \bibinfo{person}{Nicholas
  Caporusso}, {and} \bibinfo{person}{Ludiana Atnafu}.}
  \bibinfo{year}{2022}\natexlab{}.
\newblock \showarticletitle{A {Chatbot} for {Teaching} {Secure} {Programming}}.
  In \bibinfo{booktitle}{\emph{Proceedings of the EDSIG Conference ISSN}},
  Vol.~\bibinfo{volume}{2473}. \bibinfo{pages}{4901}.
\newblock


\bibitem[Wei et~al\mbox{.}(2023)]%
        {wei_chain--thought_2023}
\bibfield{author}{\bibinfo{person}{Jason Wei}, \bibinfo{person}{Xuezhi Wang},
  \bibinfo{person}{Dale Schuurmans}, \bibinfo{person}{Maarten Bosma},
  \bibinfo{person}{Brian Ichter}, \bibinfo{person}{Fei Xia},
  \bibinfo{person}{Ed Chi}, \bibinfo{person}{Quoc Le}, {and}
  \bibinfo{person}{Denny Zhou}.} \bibinfo{year}{2023}\natexlab{}.
\newblock \bibinfo{title}{Chain-of-{Thought} {Prompting} {Elicits} {Reasoning}
  in {Large} {Language} {Models}}.
\newblock
\newblock
\showeprint[arxiv]{2201.11903}~[cs.CL]


\bibitem[Weidinger et~al\mbox{.}(2022)]%
        {weidinger2022taxonomy}
\bibfield{author}{\bibinfo{person}{Laura Weidinger}, \bibinfo{person}{Jonathan
  Uesato}, \bibinfo{person}{Maribeth Rauh}, \bibinfo{person}{Conor Griffin},
  \bibinfo{person}{Po-Sen Huang}, \bibinfo{person}{John Mellor},
  \bibinfo{person}{Amelia Glaese}, \bibinfo{person}{Myra Cheng},
  \bibinfo{person}{Borja Balle}, \bibinfo{person}{Atoosa Kasirzadeh},
  \bibinfo{person}{Courtney Biles}, \bibinfo{person}{Sasha Brown},
  \bibinfo{person}{Zac Kenton}, \bibinfo{person}{Will Hawkins},
  \bibinfo{person}{Tom Stepleton}, \bibinfo{person}{Abeba Birhane},
  \bibinfo{person}{Lisa~Anne Hendricks}, \bibinfo{person}{Laura Rimell},
  \bibinfo{person}{William Isaac}, \bibinfo{person}{Julia Haas},
  \bibinfo{person}{Sean Legassick}, \bibinfo{person}{Geoffrey Irving}, {and}
  \bibinfo{person}{Iason Gabriel}.} \bibinfo{year}{2022}\natexlab{}.
\newblock \showarticletitle{Taxonomy of Risks Posed by Language Models}. In
  \bibinfo{booktitle}{\emph{2022 ACM Conference on Fairness, Accountability,
  and Transparency}} (Seoul, Republic of Korea) \emph{(\bibinfo{series}{FAccT
  '22})}. \bibinfo{publisher}{Association for Computing Machinery},
  \bibinfo{address}{New York, NY, USA}, \bibinfo{pages}{214–229}.
\newblock
\showISBNx{9781450393522}
\urldef\tempurl%
\url{https://doi.org/10.1145/3531146.3533088}
\showDOI{\tempurl}


\bibitem[Zhuo et~al\mbox{.}(2023)]%
        {zhuo2023red}
\bibfield{author}{\bibinfo{person}{Terry~Yue Zhuo}, \bibinfo{person}{Yujin
  Huang}, \bibinfo{person}{Chunyang Chen}, {and} \bibinfo{person}{Zhenchang
  Xing}.} \bibinfo{year}{2023}\natexlab{}.
\newblock \bibinfo{title}{Red teaming ChatGPT via Jailbreaking: Bias,
  Robustness, Reliability and Toxicity}.
\newblock
\newblock
\showeprint[arxiv]{2301.12867}~[cs.CL]


\end{thebibliography}

\end{document}